\documentclass[aip,amsmath,amssymb,reprint]{revtex4-2}

\usepackage{chemformula} 
\usepackage[T1]{fontenc} 
\usepackage{xcolor}
\usepackage{setspace}
\usepackage{booktabs}
\usepackage{siunitx}
\usepackage{graphicx}
\usepackage{soul}
\usepackage{dcolumn}
\usepackage{bm}
\usepackage[utf8]{inputenc}
\usepackage{mathptmx}
\usepackage{etoolbox}
\usepackage{hyperref}
\usepackage{textcomp}
\usepackage[capitalise]{cleveref}
\usepackage{upgreek}
\hypersetup{
	colorlinks,
	citecolor=blue,
	filecolor=black,
	linkcolor=blue,
	urlcolor=black
}

\makeatletter

\def\@email#1#2{%
	\endgroup
	\patchcmd{\titleblock@produce}
	{\frontmatter@RRAPformat}
	{\frontmatter@RRAPformat{\produce@RRAP{*#1\href{mailto:#2}{#2}}}\frontmatter@RRAPformat}
	{}{}
}%
\makeatother
\begin{document}
	
	
\title{Layered semiconductors integrated with polyimide thin films for high-quality valleytronic and quantum-photonic systems}

\author{Jithin T Surendran}
\affiliation{School of Physical Sciences, Indian Institute of Technology Goa, Ponda, 403401, Goa, India}

\author{Indrajeet D Prasad}
\affiliation{School of Physical Sciences, Indian Institute of Technology Goa, Ponda, 403401, Goa, India}

\author{Kenji Watanabe}
\affiliation{Research Center for Electronic and Optical Materials, National Institute for Materials
	Science,1-1 Namiki, Tsukuba 305-0044, Japan}

\author{Takashi Taniguchi}
\affiliation{Research Center for Materials Nanoarchitectonics, National Institute for Materials Science,
	1-1 Namiki, Tsukuba, 305-0044, Japan}

\author{Santosh Kumar}
\affiliation{School of Physical Sciences, Indian Institute of Technology Goa, Ponda, 403401, Goa, India}

\email{skumar@iitgoa.ac.in}
\begin{abstract}
Dielectric integration of layered semiconductors is a prerequisite for fabricating high-quality optoelectronic, valleytronic, and quantum-photonic devices. While hexagonal boron nitride (hBN) is the current benchmark dielectric, exploration of the most suitable dielectric materials covering the complete substrates continues to expand. This work demonstrates the formation of high optical-quality excitons in two widely explored layered semiconductors, WSe$_2$ and WS$_2$, integrated into polyimide (PI) thin films of thicknesses $\approx$500\,nm. The photoluminescence (PL) studies at $T$\,=\,296\,K show the formation of neutral excitons $\left(X^0\right)$ and trions in fully-PI-encapsulated 1L-WSe$_2$ with 2-sigma ($2\sigma$) spatial-inhomogeneity of 4.5 (3.4) meV in $X^0$ emission energy (linewidth), which is $\approx$1/3rd (1/5th), respectively, that of inhomogeneity measured in fully-hBN-encapsulated 1L-WSe$_2$. A smaller $2\sigma$ of 2.1 (2.3) meV in $X^0$ emission energy (linewidth) has been shown for fully-PI-encapsulated 1L-WS$_2$. Polarization-resolved and excitation power-dependent PL measurements of PI-isolated 1L-TMDs at $T$\,=\,4\,K further reveal formations of high-quality neutral-biexcitons and negatively-charged biexcitons, with degrees of valley-polarization up to 21$\%$ under non-resonant excitation. Furthermore, the fully-PI-encapsulated 1L-WSe$_2$ also hosts single quantum emitters with narrow linewidths and high-spectral stability. This work indicates that PI thin films may serve the purpose of high-quality dielectric material for integrating the layered materials on a wafer scale.
	
\end{abstract}

\maketitle
	
Transition metal dichalcogenides (TMDs) in their monolayer (1L) and a few layers (nL) thicknesses limit exhibit strong Coulomb interaction owing to weak dielectric screening, favors the formation of stable excitons with large binding energy (0.2\,-\,1.0\,eV)\cite{tmdApplic_Muller,BE_PRL_2010}. Besides the neutral-excitons $\left(X^0\right)$, the formations of high-order excitons, including negative-excitons $\left(X^-\right)$, neutral-biexcitons $\left(XX\right)$, and negative-biexcitons $\left(XX^-\right)$, combined with exotic spin-locked valley properties of 1L-TMDs makes them attractive candidates for realizing spintronic\cite{spintronics}, valleytronic\cite{valleytronics} and exciton-transport\cite{exicton_transport} devices. Moreover, the capability to host single quantum emitters\cite{he15WSe2,srivastava15,branny17} (SQEs) and their ease of integration into photonic structures promises potential application in rapidly growing quantum technologies\cite{jelena,speQuantumcommun}.
	
Excitonic- and single-photon emission- properties of 1L-TMDs are extremely sensitive to the dielectric environments present in the van der Waals heterostructures and devices. The exfoliated flakes of thicknesses <30\,nm\cite{hbncap1,tunneling} of high-pressure and high-temperature conditions grown hexagonal boron nitride (hBN\cite{taniguchisynthesis,dean2010boron}) crystals are utilized as the most promising dielectric materials for various TMD-based heterostructures and devices. The small dimensions ($\approx$500\,\textmu m)\cite{Au_assisted} of these hBN flakes are utilized for the purposes of gate insulators\cite{gateDielectric1,dielectricsRevw}, charge-carrier tunneling barriers\cite{tunneling,schotkybarrier}, and encapsulating/protective layers\cite{hBN_applications_properties}, that are posing a major challenge in producing the large numbers of circuit and devices on a wafer-scale. The epitaxial growth of hBN demands specific template substrates and complex two-stage process\cite{twostage,twostage2}. At the same time, CVD-grown hBN lacks purity\cite{hBN_applications_properties,hbn_leak1,hbn_leak2} compared to mechanically exfoliated hBN flakes\cite{dielectricsRevw}.
	
Utilization of crystalline dielectrics other than hBN such as SiO$_2$\cite{SiO2_Al2O3_effects}, Al$_2$O$_3$\cite{SiO2_Al2O3_effects,Al2O3}, and perovskites\cite{perovskite,perovskite2} have lead to the degradation in the optical quality of 1L-TMDs. Recent studies have shown that some techniques of SQE generation in 1L-TMDs also demand mechanical flexibility and plastic deformation properties of the dielectric materials. For instance, the AFM nanoindentation technique enables deterministic creation of SQEs in 1L-TMD placed on a polymethyl methacrylate (PMMA) thin film\cite{pmma1,pmma4,spe_purity}. Additionally, TMD deposited on patterned Polystyrene (PS) nano-bulges exhibited stable SQEs \cite{PS}. Among various dielectric polymers, polyimide (PI) is notable for its excellent dielectric properties\cite{PI_dielctric}, chemical resistance\cite{PI_chem}, and more importantly, excellent structural stability at cryogenic temperatures up to 4\,K\cite{PI_helium1,PI_helium2}.

In this work, we demonstrate that the utilization of PI thin film as a dielectric and$/$or encapsulating material substantially improves the optical properties of excitons, multiexcitons, and charged excitons that are formed in 1L-TMDs. Two types of PI-based van der Waals heterostructures, in particular, 1L-WSe$_2$ and 1L-WS$_2$ as the active materials, were characterized at both room-temperature (296\,K) and low-temperature (4\,K). To prepare PI-based 1L-TMD heterostructures, we fabricated PI thin films via spin-coating PI precursor solution (Polyamic acid) on a Si/SiO$_2$/Au substrate followed by thermal imidization process at an elevated temperature (see Methods for details of fabrication). We then mechanically exfoliated TMD crystals, which were transferred to the desired location via the all-dry transfer method\cite{castellanos14alldry}. The degree of imidization of PI thin films was monitored using FTIR spectroscopy (see Fig.\,S1 of Supporting Information).

As fully-hBN-encapsulated 1L-TMDs heterostructures are the current benchmarks of providing the highest optical-quality delocalized neutral-excitons $\left(X^0\right)$, charged-excitons $\left(X^{n\pm}, n\geq 1\right)$, biexcitons $\left(XX\right)$, and multi-excitons, we compared the PL characteristics of fully PI-encapsulated 1L-TMDs with those of fully hBN-encapsulated 1L-TMDs. We observed that PI-encapsulation reduces spectral- and spatial- inhomogeneities in 1L-TMDs compared to the hBN-encapsulated samples. Furthermore,  Raman spectroscopy measurements indicated negligible strain in PI-encapsulated 1L-TMDs. At 4\,K, high-order excitonic complexes were revealed, which reflects the excellent dielectric quality of PI thin film. Notably, to the best of the authors' knowledge, observation of such excitonic complexes has not been reported on polymer-based TMD heterostructures. In addition, 2L- and 3L-WSe$_2$ flakes showed remarkably enhanced excitonic emission. The polarization-dependent PL measurements demonstrated preserved valley-locking properties in 1L-WS$_2$ under a dielectric PI environment.  Further, we demonstrate localized SQEs in 1L-TMDs with sharp linewidth and high spectral stability.

\section{Results and Discussion}
\noindent\textbf{High-quality excitons with ultra-low spatial- and spectral- inhomogeneities in fully-PI-encapsulated 1L-TMDs:} Firstly, we characterized the fully-PI-encapsulated heterostructures containing 1L-WSe$_2$ (PI/1L-WSe$_2$/PI, Sample PIS1) and 1L-WS$_2$ (PI/1L-WS$_2$/PI, Sample PIS2) as light-emitting layers. We also fabricated and characterized the control sample\,1 (hBN/1L-WSe$_2$/hBN, CS1) with the aim of presenting one-to-one comparisons of the excitonic properties in the PI-encapsulated 1L-WSe$_2$ with respect to the hBN-encapsulated 1L-WSe$_2$. As a few layers-thick hBN works very well for a high-k tunneling barrier\cite{tunneling}, we also covered a portion of the 1L-WSe$_2$ flake with a 5-6 layers-thick hBN flake before encapsulating it with the top-PI layer in Sample PIS1.
\begin{figure*}
\includegraphics[width=\textwidth]{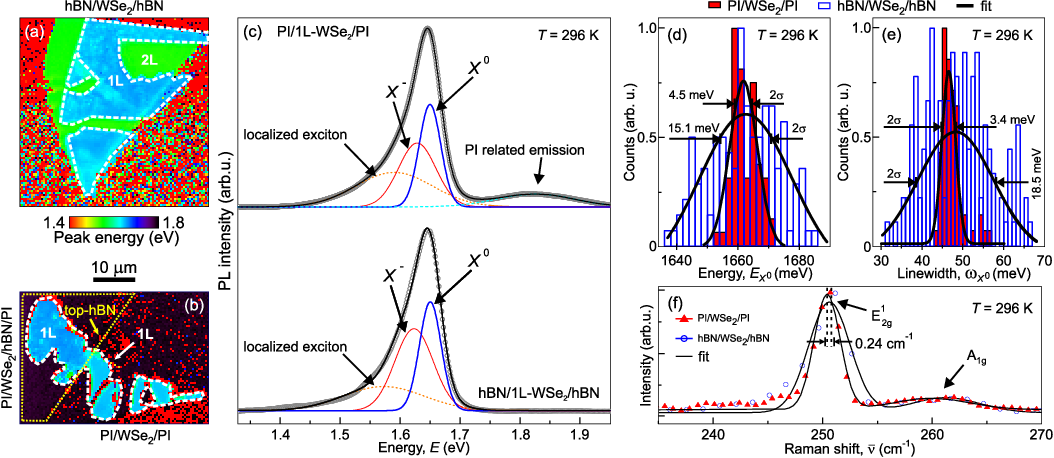}
\caption{\textbf{High-uniformity in PL properties of excitons in a fully-PI-encapsulated 1L-WSe$_2$ heterostructure at 296\,K.} \textmu -PL peak-energy spatial map of (a) Fully-hBN-encapsulated and (b) Fully-PI-encapsulated 1L-WSe$_2$ flakes. The thick dotted curve in (a) and (b) is the 1L-WSe$_2$ region. Thin dotted curve in (b) separates the left-side region where thin hBN is deposited on top of 1L-WSe$_2$ prior to the deposition of the top-PI thin film. The color codes correspond to PL peak energies in the range 1.4-1.8\,eV. Scale bar for (a) and (b): 10\,\textmu m. (c) PL spectra of 1L-WSe\textsubscript{2} under fully-PI-encapsulated (top panel) and fully-hBN-encapsulated (bottom panel) dielectric environments. The thick and thin curves are the deconvoluted peaks for delocalized exciton $\left(X^0\right)$ and trion $\left(X^-\right)$ , respectively, whereas the dotted curve is for emission due to localized excitons. Histograms of fully-hBN(PI)-encapsulated 1L WSe$_2$: (d) Peak energy and (e) Linewidth of $\left(X^0\right)$. Empty (filled) pillars correspond to fully-hBN(PI)-encapsulated 1L-WSe$_2$. Bin size in (d) and (e): 1\,meV. Bin samples for empty (filled) pillar: 66 (130). (f) Raman spectra of fully-encapsulated 1L-WSe$_2$. The measured data with open (closed) circles (triangles) are for fully-hBN(PI)-encapsulated samples. Solid lines are multi-Gaussian fits.}
\label{Fig:1}
\end{figure*}
	
Figure\,\ref{Fig:1}a and \ref{Fig:1}b shows \textmu -PL spatial maps of the peak emission energies of the WSe$_2$ flakes in the 1.4\,-\,1.8\,eV range from the h-BN encapsulated Sample CS1, and the PI-encapsulated Sample PIS1. The PL spectra for both samples were acquired at room temperature (RT) under a continuous-wave (CW) laser excitation at $\lambda$\,=532\,nm with a power of 100\,\textmu W. The representative spectra from each sample are depicted in Fig.\,\ref{Fig:1}c. These spectra were fitted using multi-Gaussian peaks for deconvoluting the various emission peaks. The thick-, thin-, dotted- curves represent the emission peaks due to formations of delocalized neutral-exciton $\left(X^0\right)$, delocalized trions $\left(X^-\right)$, and localized excitons. The detailed investigation of these localized excitons at a 4\,K temperature, leading to the emission of single photons, is presented in the later section of this work. All the PI-based samples have shown broadband weak-emission-peaks (dashed curve in Fig.\,\ref{Fig:1}c) centered at 1.820\,eV. Although the flakes of 1L-WSe$_2$ for fully-PI-encapsulated Sample PIS1 are smaller in dimensions, it can clearly be seen in Fig.\,\ref{Fig:1}b that they show similar peak emission energies through a single color contrast. On the other hand, a noticeable variation in the peak energies of PL emission of 1L-WSe$_2$ can be seen in Fig.\,\ref{Fig:1}a in the fully-hBN-encapsulated Sample CS1.
	
For understanding the spatial inhomogeneity in PL signals from these two samples further, we analyzed the statistical distribution of peak emission energies and emission linewidths of PL spectra obtained from many different locations on both the samples. The PL spectra were collected from $\approx$21\,\textmu m$^2$ area of Sample CS1, and from $\approx$11\,\textmu m$^2$ area of Sample PIS1, by avoiding the edges of the flake where abrupt changes in the emission properties are expected. Figure\,\ref{Fig:1}d shows peak energy histograms of delocalized $X^0$ of Sample CS1 (filled pillars) and Sample PIS1 (empty pillars). The solid thick lines are the Gaussian fits of the bin heights of the histograms. To our surprise, ensemble emissions of $\left(X^0\right)$ for both types of encapsulation lead to the same values at 1.651\,eV that are consistent with the literature\cite{hbncap1}, and signifying the fact that PI can be seen as a suitable alternative dielectric at the place of hBN. Interestingly, in comparison to the fully-hBN-encapsulated Sample CS1, the fully-PI-encapsulated Sample PIS1 shows a narrow distribution in the $\left(X^0\right)$ peak-energies.
	
We quantify these statistical distributions in terms of degrees of inhomogeneous broadening in $\left(X^0\right)$ emission energy, 2$\sigma_{E_{X0}}$, that are the widths of the peak-energy histograms shown in Fig.\,\ref{Fig:1}d. We obtained 2$\sigma_{E_{X0}}$\,=\,15.1\,meV for Sample CS1, while a significantly reduced value of 2$\sigma_{E_{X0}}$\,=\,4.5\,meV for Sample PIS1. We attribute higher 2$\sigma_{E_{X0}}$ of Sample CS1 is possible because thin hBN is prone to create wrinkles and bubbles at the interface between the flakes as reported previously\cite{hBNwrinkles} that may cause local strains. Alternatively, we explain 15.1\,meV of energy distribution correspond to considerable amount of 0.10\,$\%$ strain (taking 153\,meV$/\%$\,strain as the gauge factor\cite{PL_and_Raman_guage}) in Sample CS1. At the same time, reduced PL  inhomogeneity in Sample PIS1 can be explained by the uniform clean interface provided by PI thin film over the complete substrate, which shows potential for realizing valleytronic and quantum photonic devices on PI platform. Besides the $X^0$ energy, a similar observation was noted in the statistics of linewidths of $\left(X^0\right)$ emissions. Figure\,\ref{Fig:1}e shows $\left(X^0\right)$ linewidth histograms of both the Samples CS1 and PIS1. The obtained degree of inhomogeneous broadening in $\left(X^0\right)$ linewidths in Sample PIS1 $\left(2\sigma_{\omega_{X0}}\,=\,3.4\,meV\right)$ was much smaller than observed in Sample CS1 $\left(2\sigma_{\omega_{X0}}\,=\,18.5\,meV\right)$. This further confirms the formation of high-optical quality excitons throughout the fully-PI-encapsulated 1L-WSe$_2$ flake.
	
The substrate and/or encapsulation layers play important roles in determining the static doping of the TMD materials. Both qualitative and quantitative information on such doping can be obtained by monitoring the emission intensities of the trions-related emission peaks. The trion spectral weight (defined as $R_{X^-}\,=\,A_{X^-}/(A_{X^-}+A_{X^0})$, where $A_{X^{0/-}}$ is the integrated peak intensity of $X^0$/$X^-$) obtained for both the Samples CS1 and PIS1 are $\approx$\,0.54, indicating to a conclusion that the dielectric quality of PI is comparable with hBN. To investigate the changes in the crystal structure of 1L-WSe$_2$, we measured room-temperature \textmu -Raman spectra from both Samples CS1 and PIS1, as shown in Fig.\,\ref{Fig:1}f, using a laser emitting at $\lambda$\,=\,532\,nm. The fully-hBN-encapsulated Sample CS1 displays a dominant peaks at 250.64\,cm$^{-1}$ corresponding to the E$_{2g}^1$ Raman mode. Taking 250\,cm$^{-1}$ as a reference E$_{2g}^1$ peak for suspended (unstrained) 1L-WSe$_2$\cite{raman13}, a blueshift of 0.64\,cm$^{-1}$ in E$_{2g}^1$ Raman mode, that corresponds to a 0.15\,\% compressive strain\cite{PL_and_Raman_guage}, was observed in the fully-hBN-encapsulated Sample CS1. Whereas the fully-PI-encapsulated Sample PIS1 showed a smaller blueshift of 0.40\,cm$^{-1}$ and hence a reduced compressive strain of only 0.09\,\% in the 1L-WSe$_2$ that is further confirming high crystallinity of PI encapsulated 1L-WSe$_2$.
	
\begin{figure}
\includegraphics{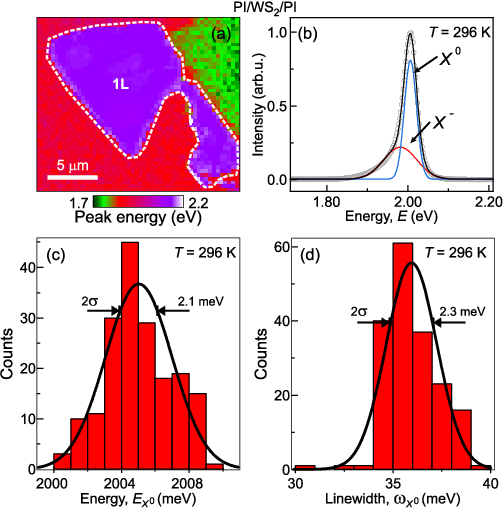}
\caption{\textbf{High-uniformity in PL properties of excitons in a fully-PI-encapsulated 1L-WS$_2$ heterostructure at 296 K.} (a) \textmu -PL peak-energy spatial map of Sample PIS2 (fully-PI-encapsulated 1L-WS$_2$ region enclosed by thick dotted curve). (b) A representative PL spectrum showing contributions from $X^0$ and $X^-$ emission lines. The open circles are the measured data, and solid lines are the multi-Gaussian fit. Histograms of (c) delocalized $X^0$ energy and (d) Linewidth of 1L-WS$_2$ emission. Bin size in (c) and (d): 1\,meV. Bin samples in (c) and (d): 181.}
\label{Fig:2}
\end{figure}

	To demonstrate that the PI encapsulation is suitable for other widely used 1L-TMD materials, we fabricated a fully-PI-encapsulated 1L-WS$_2$ Sample PIS2 (PI/1L-WS$_2$/PI heterostructure). Figure\,\ref{Fig:2}a shows room temperature \textmu -PL spatial maps of the Sample PIS2 containing a large dimensions flake of 1L-WS$_2$. Figure\,\ref{Fig:2}b displays a typical PL spectrum from the map. Fit results shows delocalized $X^0$ and $X^-$ peaks are located at 2.006\,eV (FWHM=35.6\,meV) and 1.981\,meV (FWHM=95.9\,meV) with trion spectral weight, $R_{X^-}$=0.42. For the PL statistical analysis, spectra were collected from a nearly three times bigger area than the Sample PIS1, $\approx$31\,\textmu m$^2$. The $X^0$ energy histogram (see Fig.\,\ref{Fig:2}c) reveals reduced inhomogeneity broadening with 2$\sigma_{E_{X0}}$\,=\,2.1\,meV. Similarly, $X^0$ linewidth histogram of fully-PI-encapsulated 1L-WS$_2$ (see Fig.\,\ref{Fig:2}d) also reflects reduced inhomogeneous broadening with 2$\sigma_{\omega_{X0}}$\,=\,2.3\,meV. This further proves that PI encapsulation provides a nearly uniform dielectric environment irrespective of the choice of active TMD materials.
	
\begin{figure}
\centering
\includegraphics{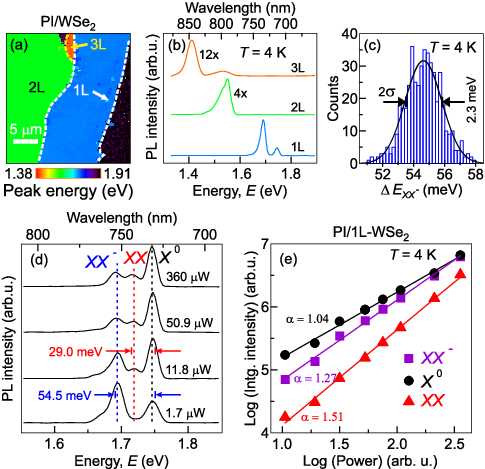}
\caption{\textbf{Excitonic complexes in PI-based 1L-WSe$_2$ at 4\,K.} (a) \textmu -PL peak-energy spatial map of Sample PIS3 (PI/WSe$_2$) containing 1L-, 2L-, and 3L-WSe$_2$ flakes marked by dotted curves. (b) PL spectra of 1L-, 2L-, and 3L WSe$_2$ from Sample PIS3 indicate strong emission from 2L and 3L regions. (c)  Histogram of binding energies ($\Delta E_{XX^-}$) of the negative biexcitons ($XX^-$) showing a narrow 2-sigma ($\sigma$)-spreading of 2.3\,meV. Bin size: 0.2\,meV. Bin samples: 471. (d)  PL spectra from 1L-WSe$_2$ of Sample PIS3 measured at various excitation powers showing the emergence of biexitons ($XX$) at and above 11.8\,\textmu W. (e) PL intensity as a function of excitation power for different excitonic species: $X^0$ (circle), $XX$ (triangle), and $XX^-$ (rectangle). The solid lines are the power-law fits, indicating linear behavior for $X^0$ and superlinear behavior for $XX$ and $XX^-$.}
\label{Fig:3}
\end{figure}

\begin{figure}
\includegraphics{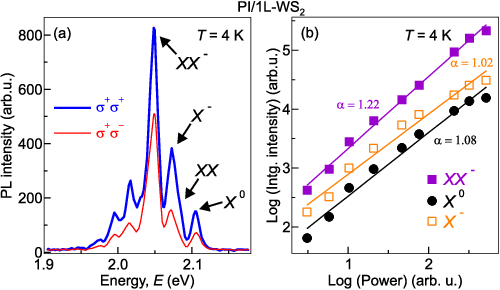}
\caption{\textbf{Valley polarization of excitons in PI-based 1L-WS\textsubscript{2} at 4\,K.} (a) Polarization-resolved PL spectra of Sample PIS4 (PI/1L-WS$_2$) showing significant valley-polarization locking. The spectrum with thick (thin) curve is for co-polarized, $\sigma^+\sigma^+$ (cross-polarized, $\sigma^+\sigma^-$) polarization configuration. (b) Excitation power-dependent PL integrated intensities of different excitonic complexes: $X^0$ (filled circle), $X^-$ (empty rectangle), and $XX^-$ (filled rectangle). The solid lines represent power-law fits, indicating linear behavior for $X^0$ and $X^-$, and superlinear behavior for $XX^-$.}
\label{Fig:4}
\end{figure}

\noindent\textbf{High-order excitonic complexes in PI-based 1L-TMD heterostructures:} For inspecting high-order excitons, we prepared Samples PIS3 and PIS4, where 1L-WSe$_2$ and 1L-WS$_2$ were isolated from substrates via PI thin films, respectively. Figure\,\ref{Fig:3}a shows low-temperature (4\,K) \textmu -PL spatial map of Sample PIS3. The flake of WSe$_2$ consists of 1L, 2L, and 3L-regions, whose representative spectra are shown in Fig.\,\ref{Fig:3}b. At low excitation power, 1L-WSe$_2$ PL has a peak at 1.750\,eV that is identified as delocalized $X^0$ and another peak at 1.695\,eV assigned to negative biexciton\cite{biexciton2} ($XX^-$). The $X^0$ linewidth is as low as 10\,meV, close to the observed value in our fully-hBN-encapsulated Sample CS1. Notably, the PL intensities of the 2L and 3L flakes are only decreased by factors of 4 and 12, respectively, indicating high exciton formation rates. The $x^0$ peak energy for 2L (3L) were observed at 1.541\,eV (1.410\,eV), that are consistent with previously reported values\cite{tonndorf2013photoluminescence}.
	The measured $XX^-$ binding energy $\Delta E_{XX^-}$ $\left(=E_{X^0}-E_{XX^-}\right)$ of 54.5\,meV agrees with the literature\cite{biexciton4,biexciton3,biexciton2}. Figure\,\ref{Fig:3}c shows the statistical distribution of $\Delta E_{XX^-}$ values obtained from fitted values of PL spectra taken from 1L-WSe$_2$ region of the flake marked in Fig.\,\ref{Fig:3}c (red curve). The considerably low fluctuations of $\Delta E_{XX^-}$ (=2.7\,meV), indicate a uniform dielectric environment provided by PI film at 4\,K temperature. This result further complements the reduced local inhomogeneity observed in 1L-WSe$_2$ at room-temperature, as discussed previously, indeed reflecting the structural stability of dielectric PI films at low temperatures. Figure\,\ref{Fig:3}d shows PL spectra acquired at different laser powers. Interestingly, when excitation power was increased above 11.8\,\textmu W, a new peak identified as neutral-biexciton\cite{biexciton2} (XX) appeared at 1.725\,eV. The measured neutral-biexciton binding energy $\Delta E_{XX}$ $\left(=E_{X^0}-E_{XX}\right)$ of 25.0\,meV is in close agreement with the previous reports\cite{biexciton4,biexciton3,biexciton2}. To further investigate the nature of the biexciton complexes, we studied  PL intensities of various peaks as a function of excitation power, as depicted in Fig.\,\ref{Fig:3}e. The extracted intensities of multiple peaks were fitted using the power-law function of the form $I\propto P^\alpha$, where $I$ and $P$ stand for integrated PL intensity and excitation power, respectively. For the $X^0$ peak, we obtain $\alpha=1.08$. Due to the linear behavior of $X^0$ with an increase in power, we attribute their origin is free from defect-bound sites\cite{defects}. On the other hand, $XX$ power-dependence shows a superlinear behavior\cite{biexcitonthermodynamics,biexciton1} with $\alpha=1.63$. Ideally, $XX$ exhibits $\alpha=2$. However, a reduced value of $\alpha$ in our sample is possibly due to competition between equilibrium states of excitons and biexcitons\cite{xx_idealAlpha}. A similar decrease was also observed in $\alpha$ (=1.22) for $XX^-$. In addition to 1L-WSe$_2$, we also investigated excitonic complexes in 1L-WS$_2$ based Sample PIS4 (PI/WS$_2$). In contrast to 1L-WSe$_2$, the 1L-WS$_2$ exhibited more complex excitonic features as shown in Fig.\,\ref{Fig:4}a (see also power-dependent spectra in Fig.\,S3 of Supporting Information). The characteristic peak at 2.109\,eV is the $X^0$ emission. The lower energy peaks at 2.077\,eV and  2.058\,eV are attributed to $X^-$ and $XX^-$, respectively. The binding energy of these excitonic complexes are $\Delta E_{X^-}$=30.2\,meV and $\Delta E_{XX^-}$=57\,meV agrees with the literature\cite{WS2Biexc}. However, a clear, distinct $XX$ peak was not observed in our Sample PIS4. Other peaks may arise from localized states and need further investigations. Interestingly, the power-law fit results (Fig.\,\ref{Fig:4}b) clearly indicate linear behavior of $X^0$ and $X^-$ with exponents $\alpha=1.08$ and $\alpha=1.02$, while XX$^-$ as expected shows superlinear trend with $\alpha=1.22$.

\noindent\textbf{Valley polarization of PI-based 1L-WS$_2$ heterostructure:} Valley polarizability in 1L-TMDs is an unprecedented property for manipulating the quantum states of a valley. We study the effect of PI on valley polarizability via performing polarization-resolved PL of Sample PIS4 at a temperature of 4\,K. We excited 1L-WS$_2$ using $\sigma^+$ (right-handed) circularly polarized light at $\lambda$\,=\,532\,nm and detected PL spectra in $\sigma^+$ and $\sigma^-$ (left-handed) in both polarization basis. Figure\,\ref{Fig:4}a shows valley-polarized PL spectra. The solid and dotted curves are taken in $\sigma^+\sigma^+$ and $\sigma^+\sigma^-$ basis, respectively. Interestingly, PL intensity between $\sigma^+\sigma^+$ and $\sigma^+\sigma^-$ of the peaks shows clear variation, suggesting the capability to maintain spin-locking information of 1L-WS$_2$ under a PI-dielectric environment. To quantitatively measure the valley polarization information, we define the degree of circular polarization: $P_c=\left(I_{\sigma^+\sigma^+}-I_{\sigma^+\sigma^-}\right)/\left(I_{\sigma^+\sigma^+}+I_{\sigma^+\sigma^-}\right)$
	, where $I_{\sigma^+\sigma^+}\left(I_{\sigma^+\sigma^-}\right)$ represents PL intensity with the same (opposite) helicity in the excitation and detection. The extracted $P_c$ values of 0.11, 0.21, and 0.19 for $X^0$, $X^-$, and $XX^-$, respectively, demonstrate the ability of dielectric-PI to sustain valley polarization properties in 1L-TMDs even under non-resonant excitation conditions. These $P_c$ values are comparable to those reported for hBN-encapsulated 1L-WS$_2$\cite{WS2_pol_4K_XX-}.

\begin{figure*}
\includegraphics{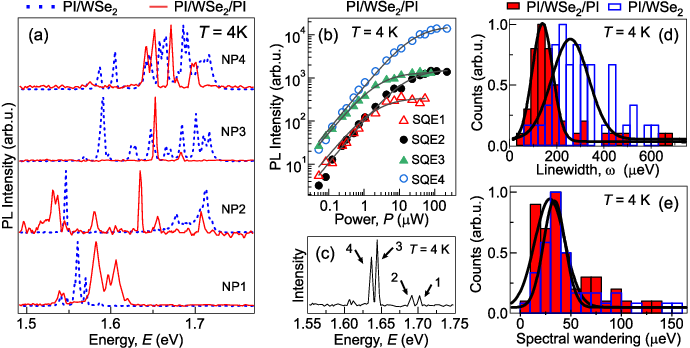}
\caption{\textbf{Single quantum emitters.} (a) PL spectra of single quantum emitters (SQEs) from multiple NP locations comparing the effect of full-PI-encapsulations. Dotted and solid curves represents PI/1L-WSe$_2$ and PI/1L-WSe$_2$/PI, respectively. (b) Excitation power dependence of the integrated intensity of SQEs from the fully-PI-encapsulated sample (PI/1L-WSe$_2$/PI). Solid curve represent fit using the relation $I=I_{sat}/\left(P_{exc}\,+\,P_{N}\right)$. $P_{N}$ is the normalization excitation power at which PL intensity becomes half of the saturation intensity $I_{sat}$. The intensity saturation behavior indicates the quantum nature of the emitters. (c) A low excitation power (125\,nW) PL spectra showing a few SQEs shown in (b). (d) Histograms of (d) Linewidth of SQEs and (e) Spectral wandering recorded for 120\,s. The empty and filled pillars represent the samples PI/1L-WSe$_2$ and PI/1L-WSe$_2$/PI. Bin size in (d) and (e): 30\,\textmu eV and 10\,\textmu eV respectively. Bin samples for empty (filled) pillars: 43 (46).}
\label{Fig:5}
\end{figure*}

\noindent\textbf{SQEs in fully-PI-encapsulated 1L-WSe$_2$ heterostructure:} Finally, we demonstrate the effect of the PI dielectric environment on optical properties of SQEs in 1L-WSe$_2$. For consistent comparison, we measured PL spectra at 4\,K for sample only with bottom PI layer (i.e., PI/1L-WSe$_2$) and fully-PI-encapsulated (i.e., PI/1L-WSe$_2$/PI) of the same WSe$_2$ flake.
	Figure\,\ref{Fig:5}a shows narrow PL spectra of SQEs from different bright spots observed in \textmu -PL map before (dotted blue curve) and after (solid red curve) full encapsulation. Such bright emission spots at nanoparticle (NP) locations (see Fig.\,S4 of Supporting Information) arise from localized excitonic states\cite{speOtica15,speLocalized19} with peak energy typically spreading in the range 1.49-1.72\,eV (see Methods for details). Both the half- and fully-encapsulated samples exhibited isolated emission lines, a key factor for improving single-photon emission purity\cite{spe_purity} by reducing multi-photon and background emission. However, some of the emitters disappeared upon full encapsulation, and new narrow peaks were formed. We ascribe this effect to the thermal treatment involved during the top-PI capping process (see Methods for details), which removes unstable emitters originating from unstable defect states in 1L-WSe$_2$\cite{NP_Surendran24}. The optical properties of these emitters were characterized by measuring their FWHM linewidths and spectral wanderings of their emission lines. Figure\,\ref{Fig:5}b compares the linewidth histogram of emitters in half- and full-encapsulated samples. Remarkably, in both the samples, we observed linewidth as low as the resolution limit (50\,\textmu eV) of our instrument.  However, the spreading of the linewidth distribution significantly reduced in the case of fully-PI-encapsulated sample. To investigate the stability of SQEs, we measured spectral-wandering, $\Delta E$, (one-standard-deviation energy from the time-averaged emission energy) as shown in Fig.\,\ref{Fig:5}c. This wandering is likely due to local fluctuations in the charge environment around the SQE site. We noticed no significant changes in spectral wandering statistics in both the samples (Fig.\,\ref{Fig:5}c), signifying no adverse effects of polymeric dielectric environment. Finally, to verify the quantum nature characteristics of SQEs, we measured the excitation power dependence of  PL intensity of a few emitters from fully-PI-encapsulated sample (Fig.\,\ref{Fig:5}e). The measured data was fitted for a two-level system using the relation $I=I_{sat}\,/\,(P_{exc}\,+\,P_{N})$. $P_{exc}$ is excitation power, and $P_{N}$ is the normalization excitation power at which intensity becomes half of the saturation intensity $I_{sat}$. The PL intensity saturation behavior of emitters clearly indicates quantum nature.
	
\section{Conclusion}
	\textcolor{black}{In summary, we have studied the optical quality of polyimide (PI)-encapsulated 1L-TMDs, including WSe$_2$ and WS$_2$ at both the temperatures of 296\,K and 4\,K. The high optical quality of our PI-encapsulated samples was evidenced by significantly low spatial- and spectral inhomogeneities with 2$\sigma$-statistical spreads within both the exciton-emission-energy and exciton-emission-linewidth at 296\,K as low as 2.1\,meV and 2.3\,meV, respectively. The average value of both exciton energy and trion spectral weight in PI-encapsulated 1L-WSe$_2$ was nearly identical to that in hBN-encapsulated 1L-WSe$_2$, indicating that PI offers a dielectric quality on par with the current benchmark where hBN is used. At 4\,K, the uniform and clean PI-dielectric environment enabled 1L-TMDs to exhibit high-order excitonic states, including neutral- and negative- biexcitons. Such complex excitonic states were previously observed predominantly in hBN-encapsulated TMD heterostructures. Moreover, PI encapsulation leads to enhanced excitonic emission even from 2L- and 3L-WSe$_2$. Polarization-resolved PL measurements under non-resonant CW excitation at 4\,K demonstrated that PI thin films can preserve the valley polarization property with degree-valley-polarization values of 11\,$\%$, 21\,$\%$, and 19\,$\%$ for excitons, trions, and negative biexcitons, respectively in 1L-WS$_2$. Further, we demonstrated localized single quantum emitters in PI-encapsulated 1L-WSe$_2$ exhibited resolution-limited linewidth and low spectral wandering. Our results show the capability of fabricating and obtaining high-quality polymer-based TMD heterostructures towards the development of room-temperature and low-temperature valleytronic and quantum photonic devices.}

\section*{Methods}
\subsection*{Sample preparation}
	PI thin films were prepared via spin coating  of commercially available polyamic acid (PPA) precursor solution (Sigma-Aldrich:431176) on a Si/SiO$_2$/Au\,(100\,nm) chip. For the uniform dispersion of PAA, a two-step spin coating was followed: first at 500 rpms$^{-1}$ for 2\,s, followed by 4000\,rpms$^{-1}$ for 60\,s. The film was then thermally cured was at 250$^{\circ}$C for 1\,hour to complete the imidization process. In order to achieve high-quality PI films, the spin-coating and thermal curing steps were carried out inside glove-box (Ar atmosphere). Thickness of PI films ($\approx$500\,nm) were measured using ellipsometry method.
	
	The TMDs (WSe$_2$ and WS$_2$) and hBN flakes were mechanically exfoliated from their bulk counterparts. The 1L-TMDs were qualitatively identified by optical color contrast and further confirmed by Raman spectroscopy. We employed the conventional dry-transfer method using polydimethylsiloxane (PDMS)  stamp for transfer the flakes on a PI thin film. To deposit top-PI layer a few drops of PPA precursor solution was dispensed onto the PI/1L-TMD, followed by spin-coating for 60\,s at 4000\,rpms$^{-1}$, and subsequently thermally cured at 250$^{\circ}$C for 1\,hour. To produce localized SQEs, Silica NPs of average diameter 150\,nm purchased from Sigma-Aldrich were spin-coated (8000\,rpm) onto the bottom-PI layer. These NPs act as nano stressors on 1L-TMDs as demonstrated in our previous work\cite{NP_Surendran24}.

\subsection*{Experimental setup}
	Photoluminescence and Raman measurements were performed using a home-built confocal microscope. The samples were mounted on an XYZ nanopositioner in a closed-cycle cryogen-free cryostat (Attodry800 from attocube systems AG) at $T$\,=\,4\,K equipped with an LT APO objective (NA\,=\,0.82). A diode-pumped solid-state CW laser emitting at $\lambda$\,=\,532\,nm was used for both PL and Raman measurements. An ultrasteep longpass filter designed at an edge of 533.3\,nm was used to block the reflected laser. All the spectra were acquired with a 0.5\,m focal length spectrometer and water-cooled charge-coupled device providing a best spectral resolution of 50\,\textmu eV (125\,\textmu eV) at $\lambda$\,=\,785\,nm ($\lambda$\,=\,532\,nm) on an 1800 lines/mm grating. Polarization-dependent PL measurements were performed at 4\,K under non-resonant CW excitation condition. Polarization basis controlled via a combination of polarizer, half wave plate, and quarter wave plate.

\section*{Author Information}
\subsection*{Corresponding Author}
*E-mail (S.K.): skumar@iitgoa.ac.in

\section*{Acknowledgement}
We thank E. S. Kannan, S. R. Parne, and A. Rahman for the fruitful discussion and A. Rastelli for the data analysis software. This work was supported by the DST Nano Mission grant (DST/NM/TUE/QM-2/2019) and the matching grant from IIT Goa. We also thank CoE: PCI, IIT Goa for access to Raman and FTIR spectroscopy facilities. I.D.P. thanks The Council of Scientific $\&$ Industrial Research (CSIR), New Delhi, for the doctoral fellowship. K.W. and T.T. acknowledge support from the JSPS KAKENHI (Grant Numbers 21H05233 and 23H02052) and World Premier International Research Center Initiative (WPI), MEXT, Japan.

\bibliography{PI_Main}
\section*{Supporting Information}
\renewcommand{\thefigure}{S\arabic{figure}}
\setcounter{figure}{0}
\section{Sample fabrication and methods}
\renewcommand{\thefigure}{S\arabic{figure}}
\textbf{Preparation of Polyimide thin film:} To fabricate polyimide (PI) thin films, we utilize commercially available polyimide precursor solution polyamic acid (PAA) purchased from Sigma Aldrich. The fabrication consists of mainly two steps:  (1) spin-coating of PAA on a Si/SiO$_2$\,(270\,nm)/Au\,(100\,nm) substrate (2) Thermal treatment of PAA (known as imidization process). The spin-coating parameters used to prepare the sample described in the main text are: spin speed of 4000\,rpm for 60\,s with two-step acceleration, first, 500\,rpm\,s$^{-1}$ for 2\,s followed by 2000\,rpm\,s$^{-1}$ for 2\,s. Immediately after spin-coating, the substrate was placed on a hot plate and thermally cured at 120$^{\circ}$C for 2\,minutes followed by heating at an elevated temperature (referred to as imidization temperature) of 250$^{\circ}$C for 45\,minutes. During the imidization process, the PAA is converted into PI form. The structural quality of PI thin films was measured by performing FT-IR spectroscopy. A comparison illustrating changes in spectra is shown in Figure\,\ref{FigS1}. The peaks at 780, 1385, and 1500\,cm$^{-1}$ correspond to out-of-plane deformation of the imide ring, C-N stretching of the imide bond, and stretching of the benzene ring, respectively. The conversion of PAA to PI film was determined by calculating the degree-of-imidization ($DoI$) expressed as\cite{DoI},
\begin{equation}
	DoI = \frac{(A_{1385}/A_{1500})_{sample}}{(A_{1385}/A_{1500})_{reference\,\,sample}} \times 100
\end{equation}
where $A$ represents the integrated area of the peak, and the reference sample corresponds to PI film prepared at 300\,$^{\circ}$C imidization temperature with 60\,minutes of imidization time. The $DoI$ values for samples prepared at  120\,$^{\circ}$C and  250\,$^{\circ}$C are 66$\%$ and 99$\%$ respectively.

\textbf{Preparation of heterostructures:} The WSe$_2$ and WS$_2$ bulk crystals were purchased from 2D Semiconductors (Scottsdale, USA). Monolayer (1L) flakes were exfoliated from their bulk counterparts via the Scotch tape method. The 1L flakes were qualitatively identified by the optical color contrast method and further confirmed by photoluminescence and Raman spectroscopy. The flakes were then deposited on the desired substrates via the all-dry transfer method. Figure\,\ref{Fig:S2} shows optical images of some of the samples mentioned in the main text.

\begin{figure} [h!]
	\includegraphics[scale=1]{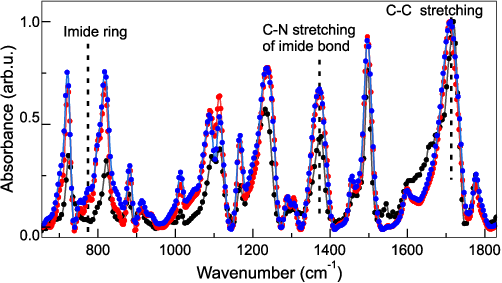}
	\caption{ FT-IR spectra of polyimide thin films prepared at different imidization temperatures. The peaks at 780, 1385, and 1500\,cm$^{-1}$ corresponds to out-of-plane deformation of the imide ring, C-N stretching of the imide bond, and C-C stretching of the benzene ring. The spectra were normalized with respect to the peak intensity of C-C stretching of each sample.}
	\label{FigS1}
\end{figure}
\begin{figure} [h!]
	\includegraphics[scale=1]{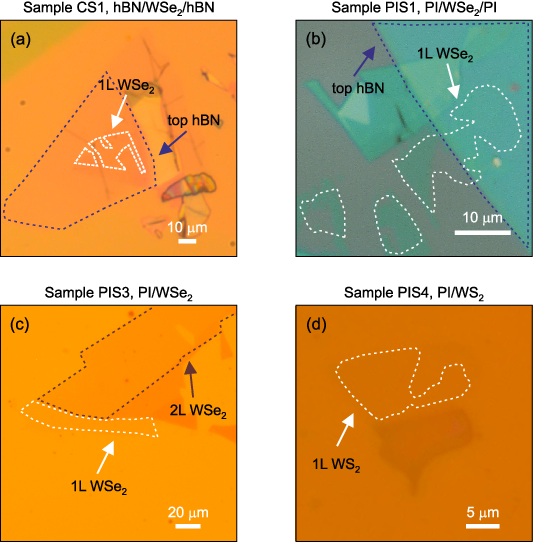}
	\caption{Optical image of Samples (a) CS1 (b) PIS1 (c) PIS3 (d) PIS4.}
	\label{Fig:S2}
\end{figure}

\newpage
\section{High-order excitons in 1L-WS$_2$}
\begin{figure} [h!]
	\includegraphics[scale=1]{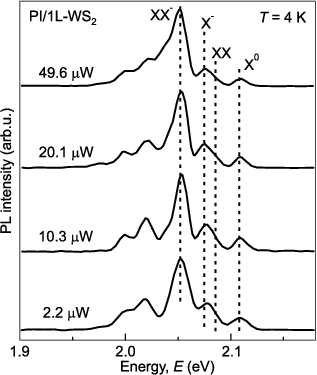}
	\caption{PL spectra from 1L-WS$_2$ of Sample PIS3 measured at various excitation powers.}
	\label{Fig:S3}
\end{figure}

\end{document}